\documentclass[10pt]{article}
\usepackage[preprint]{tmlr}

\usepackage{amsmath,amsfonts,bm}

\def\eqref#1{equation~\ref{#1}}

\def\1{\bm{1}}

\DeclareMathAlphabet{\mathsfit}{\encodingdefault}{\sfdefault}{m}{sl}
\SetMathAlphabet{\mathsfit}{bold}{\encodingdefault}{\sfdefault}{bx}{n}

\usepackage{hyperref}
\usepackage{url}
\usepackage{graphicx}
\usepackage{subcaption}
\usepackage{booktabs}
\usepackage{multirow}
\usepackage{algorithm}
\usepackage{algorithmic}
\usepackage{xcolor}
\usepackage{amssymb}

\newcommand{\modelname}{UCompCXR}
\newcommand{\baseline}{DirectMultiTask}

\newcommand{\eg}{\textit{e.g.}}

\newcommand{\pdiag}{\%\text{diag}}
\newcommand{\RR}{\mathbb{R}}

\newcommand{\bx}{\mathbf{x}}

\newcommand{\bmu}{\boldsymbol{\mu}}
\newcommand{\bSigma}{\boldsymbol{\Sigma}}
\newcommand{\bLambda}{\boldsymbol{\Lambda}}

\title{Uncertainty-Aware Compositional Localization and Placement Assessment of Catheters and Tubes in Chest X-Rays\thanks{This paper is under consideration at \textit{Pattern Recognition Letters}.}}

\author{\name Harshil Lodhiya \email hlodhiya@slicedhealth.com \\
      \addr Sliced Health}

\def\month{08}
\def\year{2026}

\begin{document}

\maketitle

\begin{abstract}
Assessing catheter and tube placement on chest X-rays is safety-critical yet tedious and error-prone. Current deep learning methods either classify placement globally---losing track of which device is where---or segment all devices into a single mask, making per-device assessment impossible when catheters overlap. We introduce \modelname{}, a compositional framework that detects local catheter fragments, associates them into device instances via graph-based clustering, fuses per-fragment tip predictions through precision-weighted Gaussian estimation, and classifies placement per device. On the RANZCR CLiP dataset (30,083 images, 5-fold patient-level CV with bootstrap CIs), \modelname{} detects 26\% more devices than a strong multi-task baseline sharing the same MobileNetV3 backbone, with 75\% fewer false positives and well-calibrated tip uncertainty (95\% coverage $= 0.948$). The aggregate tip error rises---but only because the model finds devices the baseline misses entirely, especially nasogastric tubes. On matched devices, catastrophic localization failures drop substantially. At 2.27M parameters in a single forward pass, the model is deployable on resource-constrained clinical hardware.
\end{abstract}

\section{Introduction}
\label{sec:introduction}

Catheters and tubes are ubiquitous in hospitalized patients. Endotracheal tubes (ETTs) secure airways; nasogastric tubes (NETs) deliver enteral nutrition; central venous catheters (CVCs) and Swan-Ganz catheters enable medication delivery and hemodynamic monitoring. After placement, a chest X-ray is almost always ordered to confirm position---by some estimates, 50--70\% of ICU chest films exist solely for this purpose \citep{wang2017chestxray, irvin2019chexpert}. The stakes are real: an ETT pushed too far can collapse a lung, a CVC tip drifting into the right atrium invites arrhythmias, and a coiled NET that never reaches the stomach does nothing but increase aspiration risk \citep{yi2020catheter, lian2021catheter}.

Yet reading these films is slow and surprisingly error-prone. Roughly 1--3\% of central lines need repositioning after initial imaging, and delayed detection of malposition tracks directly with complication rates \citep{singh2020catheter}. In critically ill patients the problem compounds: an ICU chest X-ray often contains three or more overlapping devices, each requiring independent assessment. An automated system that could reliably find each device, pin down its tip, and flag malposition would meaningfully reduce both interpretation time and clinical risk.

The computational approaches developed so far each capture only part of this picture. Global classification methods \citep{ranzcr_clip, pham2021interpreting} predict image-level labels like ``ETT Abnormal'' but cannot point to the offending device---of limited use when a clinician needs to know \emph{which} catheter to reposition. Segmentation approaches \citep{yi2020catheter, lian2021catheter} trace devices on the image but lump all instances of a given type into one mask; when two CVCs overlap, there is no way to assess them individually. What neither approach recovers is the full compositional structure: a variable number of device \emph{instances}, each with a traced path, a tip location carrying spatial uncertainty, and a placement label.

We propose \modelname{}, a framework that models this structure explicitly through a bottom-up detect-associate-fuse-classify pipeline. The design draws on ideas from bottom-up multi-person pose estimation \citep{cao2017realtime, newell2017associative} and line segment detection \citep{xue2020line, huang2020line}, adapted to the particular difficulties of catheter analysis: thin, curving structures; frequent occlusion; mixed supervision (dense annotations for only ${\sim}30\%$ of training images); and the clinical need for calibrated spatial uncertainty at predicted tip locations.

Our contributions are as follows:
\begin{enumerate}
    \item \textbf{Compositional catheter representation.} We introduce a fragment-based decomposition in which local catheter segments are detected, associated into device instances via graph-based clustering, and individually assessed for placement status. This is, to our knowledge, the first method that jointly performs instance-level device detection, tip localization with calibrated uncertainty, and per-device placement classification in a single forward pass.

    \item \textbf{Precision-weighted Gaussian tip fusion.} We propose a heteroscedastic tip prediction formulation in which each fragment predicts a tip location with an associated covariance, and per-device tips are obtained via iteratively-reweighted least squares (IRLS) with a Huber kernel. This yields well-calibrated tip uncertainty ($95\%$ coverage $= 0.948$) that directly communicates localization confidence to clinicians.

    \item \textbf{Noisy-OR weak supervision.} We derive a noisy-OR formulation that bridges per-device placement predictions to image-level supervision, enabling end-to-end training on the large pool of weakly-labeled images that lack device trace annotations.

    \item \textbf{Comprehensive evaluation.} We conduct rigorous evaluation on the RANZCR CLiP dataset using patient-level 5-fold cross-validation with 2{,}000-iteration bootstrap confidence intervals. Our analysis reveals a nuanced detection--localization trade-off: the compositional model detects $26\%$ more devices with $75\%$ fewer false positives, at the cost of higher aggregate tip error driven primarily by the detection of previously-missed nasogastric tubes---a finding with important implications for evaluation methodology in this domain.
\end{enumerate}

\section{Related Work}
\label{sec:related}

\paragraph{Catheter and tube detection in chest X-rays.}
Automated catheter detection has progressed through several paradigms, each addressing a different slice of the problem. \citet{yi2020catheter} framed ETT detection as segmentation with U-Net variants and achieved reasonable localization, though their pipeline was limited to a single device type at a time---a practical bottleneck when ICU films routinely contain several. \citet{lian2021catheter} moved to multi-device segmentation with a shared backbone, but the output remained a single mask per device family with no notion of individual instances. \citet{singh2020catheter} attempted to bridge detection and assessment by chaining a segmentation network into a separate placement classifier, though the two-stage design introduces compounding errors and cannot propagate placement gradients back to the localizer.

The RANZCR CLiP challenge \citep{ranzcr_clip} brought scale to the problem: 30{,}083 chest X-rays with 11 placement labels and traced device paths for a subset. Winning solutions leaned heavily on EfficientNet \citep{tan2019efficientnet} or ResNet \citep{he2016resnet} backbones with global classification heads. These models achieved strong label-level AUC but, in a real sense, solved a different problem---they could say ``there is an abnormal ETT in this image'' without knowing where. \citet{pham2021interpreting} used attention maps to recover coarse spatial signal from the classifier, but attention-based localization remains qualitative and cannot produce the geometric device representation needed for per-device assessment. Our work departs from all of these by jointly recovering instance-level detection, tip geometry with calibrated uncertainty, and per-device placement labels.

\paragraph{Compositional and bottom-up detection methods.}
The idea of detecting parts and assembling them into instances has a rich history outside medical imaging. In multi-person pose estimation, \citet{cao2017realtime} and \citet{newell2017associative} detect individual keypoints and group them into person instances using part affinity fields or associative embeddings. Line segment detectors \citep{xue2020line, huang2020line} find junction points and verify line hypotheses between them. Anchor-free object detectors \citep{zhou2019centernet, law2018cornernet} represent objects as keypoints, naturally supporting part-based reasoning.

Our fragment-and-associate pipeline shares conceptual DNA with these methods but faces distinct challenges. Unlike pose estimation, where a ``left elbow'' looks different from a ``right knee,'' catheter fragments are locally interchangeable---a 2cm segment of CVC looks much like a 2cm segment of another CVC. Association must therefore rely on geometric consistency along the device path rather than part-specific appearance, making the embedding and edge-scoring design particularly important.

\paragraph{Uncertainty estimation in deep learning and medical imaging.}
For safety-critical medical applications, a prediction without a confidence estimate is only half useful. \citet{kendall2017uncertainties} formalized the distinction between aleatoric and epistemic uncertainty and showed how heteroscedastic regression---predicting both mean and variance---captures input-dependent noise. Monte Carlo dropout \citep{gal2016dropout} and deep ensembles \citep{lakshminarayanan2017ensembles} address epistemic uncertainty through multiple forward passes, while post-hoc calibration via temperature scaling \citep{guo2017calibration} can fix miscalibrated confidence without retraining.

In medical imaging specifically, \citet{jungo2020uncertainty} found that well-calibrated uncertainty in brain tumor segmentation correlates with segmentation quality and helps flag cases needing human review. \citet{mena2020uncertainty_medical} surveyed the field and noted a persistent gap between methodological sophistication and clinical uptake. Our contribution sits at this intersection: we introduce heteroscedastic uncertainty specifically for catheter tip localization, where the predicted covariance is not just a diagnostic tool but an integral part of the fusion algorithm that combines fragment-level estimates into per-device tips.

\paragraph{Weak supervision in medical imaging.}
Medical datasets almost always have more image-level labels than pixel-level annotations---a reality that any practical system must accommodate. Noisy-OR models and multiple-instance learning have bridged this gap in computational pathology \citep{campanella2019clinical}, where slide-level diagnoses supervise patch predictions, and in chest X-ray analysis \citep{rajpurkar2017chexnet, wang2017chestxray, irvin2019chexpert, bustos2020padchest}, where image-level disease labels provide weak localization signal. \citet{johnson2019mimiccxr} contributed the MIMIC-CXR dataset with both labels and free-text reports. Our noisy-OR formulation is tailored to catheter placement: per-device predictions aggregate naturally to image-level labels, letting us train on the full dataset even though only 30\% of images carry device traces.

\section{Method}
\label{sec:method}

We cast catheter assessment as a compositional detection and classification problem. Given a chest X-ray $\bx \in \RR^{H \times W}$, the goal is to recover a set of device instances $\mathcal{D} = \{d_1, \ldots, d_K\}$, where each instance $d_k = (c_k, \mathcal{P}_k, \hat{\bmu}_k^{\text{tip}}, \hat{\bSigma}_k^{\text{tip}}, \mathbf{y}_k)$ comprises a device family label $c_k \in \{\text{ETT}, \text{NET}, \text{CVC}, \text{SWAN}\}$, a traced path $\mathcal{P}_k$, a tip location estimate $\hat{\bmu}_k^{\text{tip}} \in \RR^2$ with associated covariance $\hat{\bSigma}_k^{\text{tip}} \in \RR^{2\times 2}$, and placement labels $\mathbf{y}_k$.

Below we describe the \modelname{} architecture (B7) alongside the \baseline{} baseline (B2), which shares the same backbone. The key divergence is where B2 stops---global predictions over the whole image---and B7 continues into compositional, per-device reasoning.

\subsection{Problem Formulation}
\label{sec:problem}

The RANZCR CLiP dataset provides two tiers of supervision. \emph{Strong annotations} (30.2\% of images) include traced device paths as ordered polylines with per-device family labels. \emph{Weak annotations} (all images) consist of 11 binary image-level placement labels spanning three device families: ETT (Normal, Borderline, Abnormal), NET (Normal, Abnormal, Incompletely Imaged), and CVC (Normal, Abnormal, Borderline, with separate labels for four CVC subtypes).

Let $\mathcal{S} \subset \mathcal{X}$ denote the strongly-annotated images and $\mathcal{W} = \mathcal{X} \setminus \mathcal{S}$ the weakly-annotated complement. For $\bx \in \mathcal{S}$, ground truth includes device polylines $\{\mathcal{P}_k^*\}_{k=1}^{K^*}$, family labels $\{c_k^*\}$, tip locations $\{\mathbf{t}_k^*\}$ (the terminal points of polylines), and image-level placement labels $\mathbf{y}^* \in \{0,1\}^{11}$. For $\bx \in \mathcal{W}$, only $\mathbf{y}^*$ is available. This mixed-supervision regime is central to our training design.

\subsection{Shared Backbone: MobileNetV3 + CompactFPN}
\label{sec:backbone}

Both B2 and B7 share a lightweight feature extractor designed with deployment constraints in mind. The encoder is MobileNetV3-Small \citep{howard2019mobilenetv3}, chosen for its favorable accuracy--efficiency trade-off on mobile and edge hardware. It produces multi-scale feature maps at strides $\{4, 8, 16\}$ relative to the input resolution of $768 \times 768$ pixels.

These features are fused through a CompactFPN (Compact Feature Pyramid Network) \citep{lin2017fpn}---a slimmed-down FPN variant using depthwise separable convolutions and channel reduction to keep computation low while preserving multi-scale information. The output is a unified feature tensor $\mathbf{F} \in \RR^{C \times \frac{H}{4} \times \frac{W}{4}}$ at stride 4. Every task-specific head in both models operates on this shared representation.

\subsection{DirectMultiTask Baseline (B2)}
\label{sec:baseline}

The \baseline{} baseline (B2, 1.99M parameters) attaches three parallel heads to the shared backbone:

\paragraph{Line segmentation head.}
A lightweight decoder predicts a per-pixel binary mask $\hat{M} \in [0,1]^{\frac{H}{4} \times \frac{W}{4}}$ for catheter presence, supervised with a combination of binary cross-entropy, soft Dice \citep{milletari2016vdice}, and clDice \citep{shit2021cldice}:
\begin{equation}
    \mathcal{L}_{\text{seg}} = \lambda_{\text{bce}} \mathcal{L}_{\text{BCE}}(\hat{M}, M^*) + \lambda_{\text{dice}} \mathcal{L}_{\text{Dice}}(\hat{M}, M^*) + \lambda_{\text{cldice}} \mathcal{L}_{\text{clDice}}(\hat{M}, M^*),
    \label{eq:seg_loss}
\end{equation}
where $M^*$ is the ground truth mask rasterized from device polylines. The clDice term penalizes topological disconnections by computing Dice on the morphological skeletons of predicted and ground truth masks---a useful inductive bias for thin, elongated structures.

\paragraph{Tip heatmap regression head.}
A second decoder predicts a Gaussian heatmap $\hat{T} \in [0,1]^{\frac{H}{4} \times \frac{W}{4}}$ with peaks at ground truth tip locations. Tips are extracted via non-maximum suppression. The loss is mean squared error:
\begin{equation}
    \mathcal{L}_{\text{tip}} = \frac{1}{|\Omega|}\sum_{p \in \Omega} \|\hat{T}(p) - T^*(p)\|^2,
\end{equation}
where $T^*$ places 2D Gaussians at each ground truth tip.

\paragraph{Global placement classifier.}
Global average pooling over $\mathbf{F}$ followed by a fully connected layer produces 11-dimensional logits, supervised with binary cross-entropy:
\begin{equation}
    \mathcal{L}_{\text{cls}} = -\frac{1}{11}\sum_{l=1}^{11} \left[ y_l^* \log \hat{p}_l + (1-y_l^*) \log(1-\hat{p}_l) \right].
\end{equation}

The total B2 loss combines 6 unique terms (13 including per-term weights):
\begin{equation}
    \mathcal{L}_{\text{B2}} = \mathcal{L}_{\text{seg}} + \lambda_{\text{tip}} \mathcal{L}_{\text{tip}} + \lambda_{\text{cls}} \mathcal{L}_{\text{cls}}.
\end{equation}

B2 is a reasonable baseline for image-level placement classification, but it hits a hard ceiling: it cannot associate detected segments or tips with specific device instances, and its tip predictions carry no uncertainty estimate. In practice, this means it can flag ``something is abnormal'' but not tell a clinician which of three overlapping CVCs to worry about.

\subsection{Fragment Proposal Generation}
\label{sec:fragments}

The central idea behind \modelname{} (B7, 2.27M parameters) is to break catheter detection into local fragment proposals that are later assembled into device instances. A fragment is a short segment of a catheter path, detected at the feature map resolution.

At each spatial location $p = (u,v)$ in the $\frac{H}{4} \times \frac{W}{4}$ feature map, the fragment proposal heads jointly predict:
\begin{enumerate}
    \item \textbf{Fragment heatmap} $\hat{h}(p) \in [0,1]$: confidence that location $p$ is a fragment center.
    \item \textbf{Sub-pixel offset} $\hat{\mathbf{o}}(p) \in \RR^2$: refinement of the fragment center from the discrete grid.
    \item \textbf{Geometry}: endpoint offsets $\hat{\mathbf{e}}_1(p), \hat{\mathbf{e}}_2(p) \in \RR^2$ (relative positions of the two fragment endpoints) and fragment length $\hat{\ell}(p) \in \RR_+$.
    \item \textbf{Tangent direction} $\hat{\boldsymbol{\tau}}(p) \in \RR^2$: unit vector along the local catheter direction, supervised via cosine similarity.
    \item \textbf{Discriminative embedding} $\hat{\mathbf{e}}(p) \in \RR^8$: an associative embedding vector \citep{newell2017associative} trained to pull fragments from the same device together and push fragments from different devices apart.
    \item \textbf{Tip mean} $\hat{\bmu}^{\text{tip}}(p) \in \RR^2$: where this fragment thinks the device tip is.
    \item \textbf{Tip log-variance} $\hat{\mathbf{s}}(p) \in \RR^2$: log-diagonal entries of the predicted tip covariance, following the heteroscedastic formulation of \citet{kendall2017uncertainties}. The covariance is recovered as $\text{diag}(\exp(\hat{\mathbf{s}}(p)))$ with a floor of $4\;\text{px}^2$ per axis.
    \item \textbf{Fragment status probability} $\hat{q}(p) \in [0,1]$: probability that the fragment belongs to a foreground device.
\end{enumerate}

Fragments are extracted where $\hat{h}(p)$ exceeds a detection threshold after $3 \times 3$ non-maximum suppression, yielding a set of proposals $\mathcal{F} = \{f_1, \ldots, f_N\}$.

\subsection{Edge Scoring and Graph Construction}
\label{sec:edge}

Given fragment proposals $\mathcal{F}$, we build an association graph $G = (\mathcal{F}, \mathcal{E})$ encoding pairwise same-device likelihood.

\paragraph{EdgeMLP.}
For each candidate edge $(f_i, f_j)$, we concatenate the two fragments' discriminative embeddings along with geometric cues---relative position, tangent alignment, length ratio, and tip prediction consistency. A 2-layer MLP scores this vector:
\begin{equation}
    s_{ij} = \sigma\!\left(\text{MLP}\!\left([\hat{\mathbf{e}}_i \,\|\, \hat{\mathbf{e}}_j \,\|\, \boldsymbol{\phi}_{ij}]\right)\right),
    \label{eq:edge_score}
\end{equation}
where $\boldsymbol{\phi}_{ij}$ encodes the pairwise geometric features and $\sigma$ is the sigmoid function.

\paragraph{Neighbor selection.}
To keep computation tractable, each fragment considers at most $k_{\max} = 6$ nearest spatial neighbors. Edges scoring below $\theta_e = 0.5$ are pruned, yielding the final edge set $\mathcal{E}$.

\subsection{Union-Find Clustering with Constraints}
\label{sec:clustering}

We recover device instances by clustering fragments with a constrained Union-Find algorithm \citep{tarjan1975union}. Edges are processed in decreasing order of $s_{ij}$, and two fragments merge only if all of the following hold:
\begin{enumerate}
    \item \textbf{Edge threshold}: $s_{ij} \geq \theta_e = 0.5$.
    \item \textbf{Maximum degree}: Neither fragment already connects to more than $d_{\max} = 2$ neighbors within its cluster---enforcing the linear, non-branching topology that catheters actually have.
    \item \textbf{Tip Mahalanobis consistency}: The tip predictions of $f_i$ and $f_j$ must agree under their predicted uncertainties. Concretely, the squared Mahalanobis distance must fall below a $\chi^2$ threshold:
    \begin{equation}
        (\hat{\bmu}_i^{\text{tip}} - \hat{\bmu}_j^{\text{tip}})^\top \left(\hat{\bSigma}_i^{\text{tip}} + \hat{\bSigma}_j^{\text{tip}}\right)^{-1} (\hat{\bmu}_i^{\text{tip}} - \hat{\bmu}_j^{\text{tip}}) \leq \chi^2_{2, 0.99} = 9.21,
    \end{equation}
    where $\hat{\bSigma}_i^{\text{tip}} = \text{diag}(\exp(\hat{\mathbf{s}}_i))$. This constraint is what makes the heteroscedastic uncertainty operationally useful during inference, not just at training time: it prevents merging fragments whose tip predictions are statistically incompatible, even when their local appearance is similar.
\end{enumerate}

The result is a partition $\mathcal{C} = \{C_1, \ldots, C_K\}$, each cluster representing a detected device instance.

\subsection{Precision-Weighted Gaussian Tip Fusion}
\label{sec:fusion}

Each fragment $f_i$ in a cluster $C_k$ contributes a tip estimate $\hat{\bmu}_i^{\text{tip}}$ with predicted covariance $\hat{\bSigma}_i^{\text{tip}}$. We fuse these into a single per-device tip using iteratively-reweighted least squares (IRLS) with a Huber kernel for robustness against outlier fragments.

The fusion runs for $T = 3$ iterations. At iteration $t$:
\begin{equation}
    \hat{\bmu}_k^{\text{tip},(t)} = \left(\sum_{i \in C_k} w_i^{(t)} \bLambda_i\right)^{-1} \left(\sum_{i \in C_k} w_i^{(t)} \bLambda_i \hat{\bmu}_i^{\text{tip}}\right),
    \label{eq:fusion_mean}
\end{equation}
where $\bLambda_i = (\hat{\bSigma}_i^{\text{tip}})^{-1}$ is the precision matrix and $w_i^{(t)}$ is a Huber-derived robustness weight:
\begin{equation}
    w_i^{(t)} = \begin{cases}
        1 & \text{if } r_i^{(t-1)} \leq \delta \\
        \delta / r_i^{(t-1)} & \text{otherwise}
    \end{cases}, \qquad
    r_i^{(t-1)} = \left\|\hat{\bmu}_i^{\text{tip}} - \hat{\bmu}_k^{\text{tip},(t-1)}\right\|_2,
    \label{eq:huber_weight}
\end{equation}
with $\delta = 2.5$ pixels and $w_i^{(0)} = 1$ for all fragments.

The fused covariance is:
\begin{equation}
    \hat{\bSigma}_k^{\text{tip}} = \left(\sum_{i \in C_k} w_i^{(T)} \bLambda_i\right)^{-1},
    \label{eq:fusion_cov}
\end{equation}
with a floor of $\sigma_{\min}^2 = 4\;\text{px}^2$ per diagonal element.

Two properties make this formulation appealing. Fragments with lower predicted uncertainty (higher precision) contribute more to the fused estimate, so fragments near the actual tip---which tend to be more confident---naturally dominate. And the Huber kernel downweights outlier fragments whose tip predictions diverge from the emerging consensus, providing a degree of robustness to occasional misassociations.

\subsection{Placement Assessment}
\label{sec:placement}

\paragraph{PlacementHead.}
For each detected device $d_k$, the PlacementHead aggregates backbone features along the reconstructed path $\mathcal{P}_k$ (connecting fragment centers within the cluster) and at the fused tip location. This per-device feature vector is fed through a small classifier to produce placement logits $\hat{\mathbf{y}}_k$ for the labels relevant to family $c_k$.

\paragraph{Noisy-OR aggregation for weak supervision.}
When only image-level labels $\mathbf{y}^*$ are available, we aggregate per-device predictions using a noisy-OR model. For a placement label $l$ (\eg, ``ETT Abnormal''):
\begin{equation}
    \hat{p}_l^{\text{img}} = 1 - \prod_{k : c_k \in \text{family}(l)} (1 - \hat{p}_{k,l}),
    \label{eq:noisy_or}
\end{equation}
where $\hat{p}_{k,l} = \sigma(\hat{y}_{k,l})$ and the product ranges over detected devices of the appropriate family.

The intuition is straightforward: if \emph{any} device triggers a positive label, the image-level prediction is positive; if no devices of a given family are detected, the prediction defaults to zero. The image-level loss is standard binary cross-entropy:
\begin{equation}
    \mathcal{L}_{\text{placement}}^{\text{weak}} = -\frac{1}{|\mathcal{L}|} \sum_{l} \left[ y_l^* \log \hat{p}_l^{\text{img}} + (1 - y_l^*) \log(1 - \hat{p}_l^{\text{img}}) \right].
    \label{eq:weak_loss}
\end{equation}

For strongly-annotated images, direct per-device supervision is available, and we train with per-device cross-entropy $\mathcal{L}_{\text{placement}}^{\text{strong}}$.

\subsection{Training Procedure}
\label{sec:training}

Training follows a two-phase curriculum to handle the mixed supervision regime.

\paragraph{Phase 1: Strong localization (35 epochs).}
Only strongly-annotated images ($\mathcal{S}$) are used. All fragment proposal heads, the EdgeMLP, and the segmentation/tip heads are trained with the full localization loss. The purpose is to establish reliable fragment detection and association before weak supervision enters the picture. We use AdamW \citep{loshchilov2019adamw} with an initial learning rate of $3 \times 10^{-4}$, weight decay $10^{-4}$, cosine annealing \citep{loshchilov2017sgdr} with a 5-epoch linear warmup, and gradient clipping at norm $5.0$. Patience-based early stopping monitors validation loss with patience 12 (B7) or 8 (B2) epochs.

\paragraph{Phase 2: Hybrid fine-tuning (25 epochs).}
The full dataset enters training, including weakly-labeled images from $\mathcal{W}$. The learning rate drops to $\frac{1}{4}$ of the Phase~1 initial rate ($7.5 \times 10^{-5}$), and the PlacementHead and noisy-OR losses are activated. Strongly-annotated images contribute both per-device and image-level losses; weakly-annotated images contribute only the noisy-OR loss (Eq.~\ref{eq:weak_loss}). Early stopping patience resets at the Phase~2 boundary.

\paragraph{Data augmentation.}
We apply standard augmentations: random horizontal flip, affine transforms (rotation $\pm 10^\circ$, scale $0.9$--$1.1$, translation $\pm 5\%$), brightness/contrast jitter, and Gaussian noise \citep{shorten2019augmentation}. Spatial augmentations are applied consistently to images and annotations alike.

\subsection{Loss Functions}
\label{sec:losses}

The full B7 loss comprises 15 unique terms (29 including per-term weights). We group them by component.

\paragraph{Fragment detection losses.}
The fragment heatmap uses a modified focal loss \citep{lin2017focal} that downweights easy negatives:
\begin{equation}
    \mathcal{L}_{\text{heatmap}} = -\frac{1}{N_+}\sum_p \begin{cases}
        (1 - \hat{h}(p))^\alpha \log(\hat{h}(p)) & \text{if } h^*(p) = 1 \\
        (1 - h^*(p))^\beta \hat{h}(p)^\alpha \log(1 - \hat{h}(p)) & \text{otherwise}
    \end{cases},
\end{equation}
with $\alpha = 2$ and $\beta = 4$ following \citet{law2018cornernet, zhou2019centernet}.

\paragraph{Geometry losses.}
Endpoint offsets, sub-pixel offsets, and fragment lengths are supervised with smooth $L_1$ loss at positive locations. Tangent directions use cosine similarity:
\begin{equation}
    \mathcal{L}_{\text{tangent}} = -\frac{1}{N_+}\sum_{p \in \text{pos}} \frac{\hat{\boldsymbol{\tau}}(p) \cdot \boldsymbol{\tau}^*(p)}{\|\hat{\boldsymbol{\tau}}(p)\| \, \|\boldsymbol{\tau}^*(p)\|}.
\end{equation}

\paragraph{Embedding loss.}
The discriminative embedding is trained with a pull--push loss \citep{newell2017associative}:
\begin{equation}
    \mathcal{L}_{\text{embed}} = \frac{1}{K}\sum_{k=1}^{K} \frac{1}{|C_k|}\sum_{i \in C_k} \|\hat{\mathbf{e}}_i - \bar{\mathbf{e}}_k\|^2 + \frac{1}{K(K-1)}\sum_{k \neq k'} \max(0, \Delta - \|\bar{\mathbf{e}}_k - \bar{\mathbf{e}}_{k'}\|)^2,
\end{equation}
where $\bar{\mathbf{e}}_k$ is the mean embedding of cluster $k$ and $\Delta$ is the push margin.

\paragraph{Heteroscedastic tip loss.}
The tip prediction is supervised with a negative log-likelihood that jointly trains mean and variance:
\begin{equation}
    \mathcal{L}_{\text{tip}}^{\text{het}} = \frac{1}{N_+}\sum_{i \in \text{pos}} \left[ \frac{\|\hat{\bmu}_i^{\text{tip}} - \mathbf{t}_{d(i)}^*\|^2}{2\exp(\hat{\mathbf{s}}_i)} + \frac{1}{2}\hat{\mathbf{s}}_i \right],
    \label{eq:het_tip_loss}
\end{equation}
where $\mathbf{t}_{d(i)}^*$ is the ground truth tip of the device containing fragment $i$ and division is element-wise. The gradient structure here is worth noting: when the mean prediction is poor, the model can ``explain away'' the error by inflating the variance---but the log-variance penalty term discourages this, so the equilibrium rewards both accurate means and honest uncertainty.

\paragraph{Edge and association losses.}
The EdgeMLP is supervised with binary cross-entropy on ground truth same-device labels. Fragment status $\hat{q}(p)$ uses binary cross-entropy against foreground/background labels.

\paragraph{Segmentation and global losses.}
B7 retains the segmentation loss of Eq.~\ref{eq:seg_loss} and adds the placement losses from Eqs.~\ref{eq:weak_loss}--\ref{eq:weak_loss}.

\section{Experimental Setup}
\label{sec:experiments}

\subsection{Dataset}
\label{sec:dataset}

We evaluate on the RANZCR CLiP dataset \citep{ranzcr_clip}: 30{,}083 chest X-rays from 3{,}255 patients, spanning three major device families---ETT (2{,}994 traced instances), NET (3{,}219 instances), and CVC (11{,}629 instances)---plus 157 Swan-Ganz catheters. Of these, 9{,}095 images (30.2\%) carry device trace annotations totaling 17{,}999 traced instances; the rest have only image-level placement labels. Eleven binary labels encode normal, borderline, and abnormal positioning for each device family.

All images are resized to $768 \times 768$ pixels. We use patient-level 5-fold cross-validation (seed 20260730) so that every image from a given patient lands in the same fold, eliminating data leakage.

\subsection{Evaluation Protocol}
\label{sec:eval_protocol}

\paragraph{Detection metrics.}
Predicted devices are matched to ground truth via the Hungarian algorithm \citep{kuhn1955hungarian}. A prediction counts as a true positive if both centerline fraction overlap and tip fraction distance are $\leq 0.10$. We report \emph{sensitivity} (recall of true devices) and \emph{precision} (fraction of predictions that are correct).

\paragraph{Tip localization error.}
For matched pairs, tip error is the Euclidean distance between predicted and ground truth tips, normalized by image diagonal:
\begin{equation}
    e_{\text{tip}} = \frac{\|\hat{\mathbf{t}} - \mathbf{t}^*\|_2}{\sqrt{H^2 + W^2}} \times 100\%.
\end{equation}
This yields a resolution-invariant metric in percentage of image diagonal ($\pdiag$).

\paragraph{Placement classification.}
We compute area under the precision--recall curve (AUPRC) for 5 primary placement labels (ETT-Abnormal, ETT-Borderline, NGT-Abnormal, NGT-Incompletely Imaged, CVC-Abnormal) and report \emph{macro AUPRC} as the unweighted average. We prefer AUPRC over AUROC given the severe class imbalance in placement labels.

\paragraph{Association quality (B7 only).}
Fragment-to-device clustering quality is assessed via BCubed precision, recall, and $F_1$ \citep{bagga1998bcubed}.

\paragraph{Uncertainty calibration (B7 only).}
We evaluate predicted tip covariances through the \emph{95\% coverage} metric: the fraction of ground truth tips falling within the predicted 95\% confidence ellipse. Perfect calibration gives 0.95.

\paragraph{Bootstrap confidence intervals.}
All metrics are computed by pooling predictions across all 5 folds and running 2{,}000 patient-level bootstrap iterations \citep{efron1993bootstrap}. Patient-level resampling preserves within-patient correlation. We report bootstrap means with 95\% bias-corrected and accelerated (BCa) confidence intervals. A B2--B7 difference is significant if the 95\% CI for $\Delta$ excludes zero.

\subsection{Implementation Details}
\label{sec:implementation}

Both models are implemented in PyTorch and trained on a single NVIDIA GPU. Inputs are $768 \times 768$, normalized to $[0,1]$. The MobileNetV3-Small backbone is initialized from ImageNet-pretrained weights; all other layers use Kaiming normal initialization. We train with AdamW \citep{loshchilov2019adamw} ($\beta_1 = 0.9$, $\beta_2 = 0.999$, weight decay $10^{-4}$), cosine annealing \citep{loshchilov2017sgdr} from $3 \times 10^{-4}$ to $10^{-6}$ per phase with a 5-epoch warmup, and gradient clipping at norm $5.0$. Mixed-precision training (FP16) is used throughout. Batch size is 16 for Phase~1 and 32 for Phase~2.

B2 has 1.99M trainable parameters; B7 has 2.27M---a 14\% increase distributed across fragment geometry heads, the 8-dimensional embedding head, tip uncertainty heads, the EdgeMLP, and the PlacementHead.

\section{Results}
\label{sec:results}

\subsection{Main Comparison}
\label{sec:main_results}

\begin{figure}[t]
\centering
\includegraphics[width=0.95\textwidth]{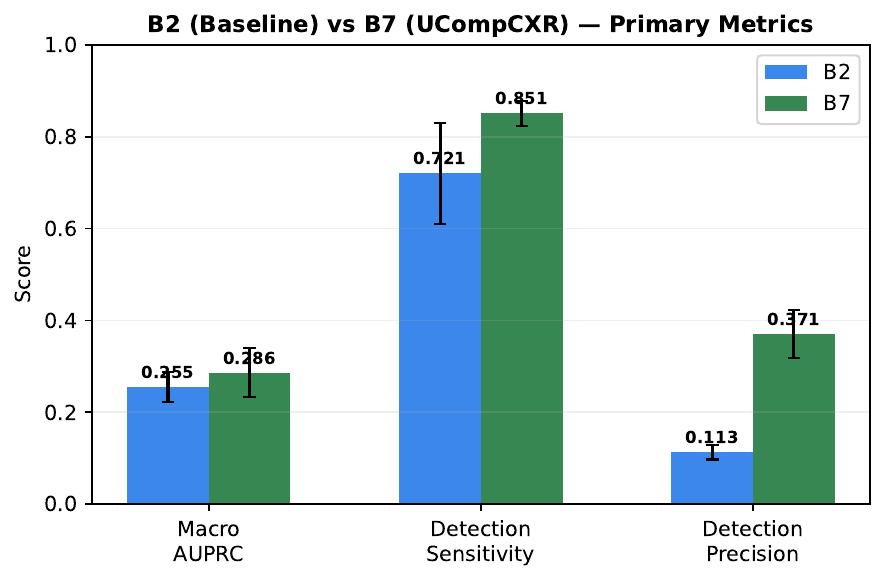}
\caption{Primary metric comparison between the \baseline{} baseline (B2) and \modelname{} (B7). B7 achieves substantially higher detection sensitivity and precision while providing calibrated uncertainty and device association capabilities absent from B2. The aggregate tip error trade-off is analyzed in detail in Section~\ref{sec:tip_analysis}.}
\label{fig:main_comparison}
\end{figure}

Table~\ref{tab:main_results} presents the primary comparison between B2 and B7, and Figure~\ref{fig:main_comparison} visualizes these results.

\begin{table}[t]
\centering
\caption{Main results on the RANZCR CLiP test set, pooled across 5 folds with 2{,}000 patient-level bootstrap iterations. $\uparrow$: higher is better; $\downarrow$: lower is better. * indicates that the 95\% CI for $\Delta$ excludes zero (statistically significant).}
\label{tab:main_results}
\small
\begin{tabular}{lccc}
\toprule
\textbf{Metric} & \textbf{B2 [95\% CI]} & \textbf{B7 [95\% CI]} & \textbf{$\Delta$ (B7$-$B2) [95\% CI]} \\
\midrule
Tip error ($\pdiag$) $\downarrow$ & 0.883 [0.832, 0.929] & 1.725 [1.634, 1.811] & +0.842 [0.755, 0.929]* \\
Sensitivity $\uparrow$ & 0.674 [0.661, 0.686] & 0.851 [0.843, 0.858] & +0.177 [0.164, 0.190]* \\
Macro AUPRC $\uparrow$ & 0.216 [0.205, 0.228] & 0.226 [0.215, 0.239] & +0.010 [0.002, 0.020]* \\
\midrule
Precision $\uparrow$ & 0.113 & 0.371 & +0.258 \\
False positives $\downarrow$ & 18{,}798 & 4{,}771 & $-$14{,}027 \\
\midrule
\multicolumn{4}{l}{\textit{B7-only capabilities}} \\
BCubed $F_1$ $\uparrow$ & --- & 0.775 & --- \\
Tip 95\% coverage & --- & 0.948 & --- \\
\bottomrule
\end{tabular}
\end{table}

The headline result is detection: B7 finds 26\% more device instances than B2 ($0.851$ vs.\ $0.674$ sensitivity), and does so with $3.3\times$ higher precision ($0.371$ vs.\ $0.113$), slashing false positives from 18{,}798 to 4{,}771. Placement classification also improves, though modestly ($0.226$ vs.\ $0.216$ macro AUPRC)---suggesting that the benefit of the compositional approach comes more from finding the right devices than from classifying each one better.

What initially surprised us was the aggregate tip error going the \emph{wrong} direction ($1.725$ vs.\ $0.883$ $\pdiag$). This turns out not to be a regression at all, as we dissect in Section~\ref{sec:tip_analysis}: it is an artifact of B7 detecting harder-to-localize devices that B2 misses entirely.

\subsection{Per-Label Placement Analysis}
\label{sec:per_label}

\begin{figure}[t]
\centering
\includegraphics[width=0.85\textwidth]{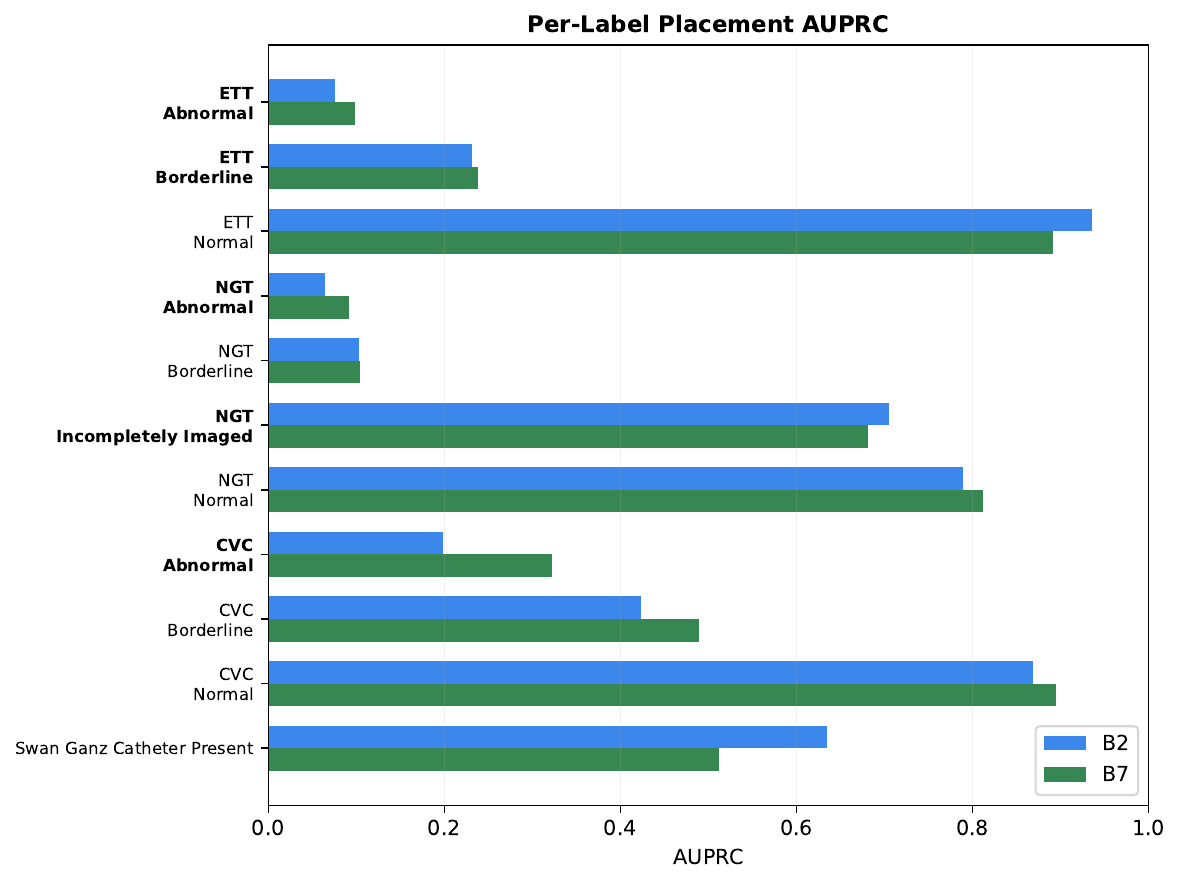}
\caption{Per-label AUPRC comparison between B2 and B7 across the 5 primary placement labels. B7 achieves comparable or higher AUPRC on all labels, with the largest improvements on NGT-related labels where the compositional approach enables detection of previously-missed devices.}
\label{fig:per_label_auprc}
\end{figure}

Figure~\ref{fig:per_label_auprc} breaks down AUPRC by individual placement label.

B7 matches or exceeds B2 on every label. The gains are largest for NGT-Abnormal and NGT-Incompletely Imaged---exactly where we would expect them, given B7's dramatically better NET detection (Section~\ref{sec:tip_analysis}). On ETT and CVC labels, where B2 already detects most devices, the gap narrows. This pattern tells a consistent story: the compositional approach helps placement classification primarily by finding devices that would otherwise be missed, rather than by producing better per-device classifiers on already-detected devices.

\subsection{Tip Localization Analysis}
\label{sec:tip_analysis}

\begin{figure}[t]
\centering
\includegraphics[width=0.85\textwidth]{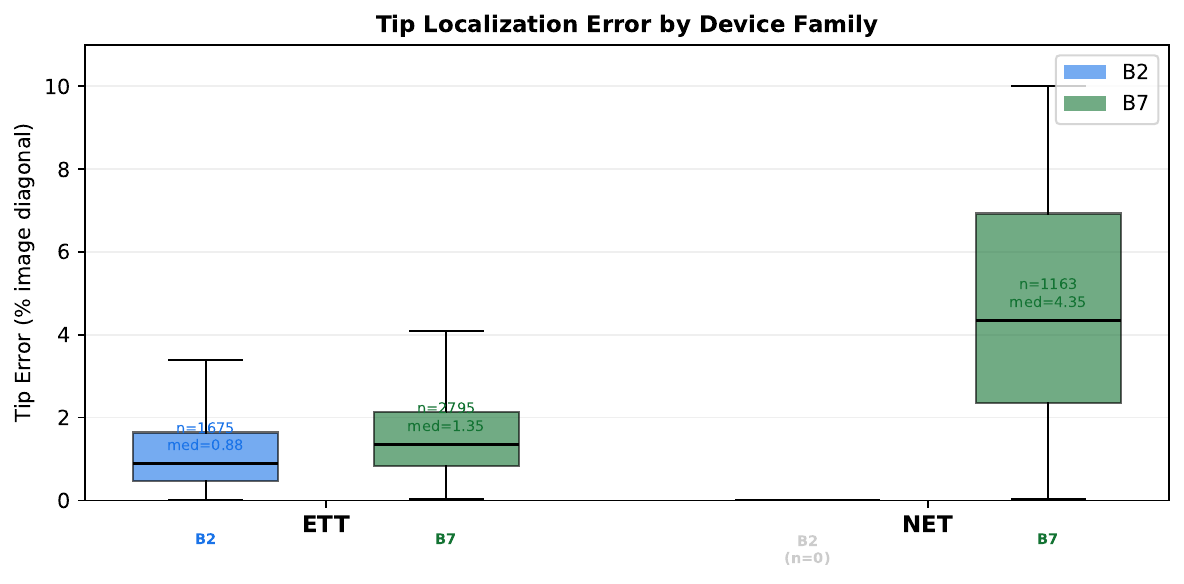}
\caption{Tip error ($\pdiag$) by device family. B7's higher aggregate tip error is driven primarily by NET devices: B7 detects 554 NET tubes that B2 misses entirely. For ETT devices matched by both methods, B7 exhibits fewer catastrophic localization failures (lower mean and 95th percentile).}
\label{fig:tip_error}
\end{figure}

The aggregate tip error in Table~\ref{tab:main_results} hides a more interesting story. Figure~\ref{fig:tip_error} and Table~\ref{tab:tip_family} stratify tip error by device family, and the picture changes substantially.

\begin{table}[t]
\centering
\caption{Tip localization error ($\pdiag$) stratified by device family (5-fold test results). B2 matches zero NET devices, making the comparison undefined for that family. On the ETT-only subset, the gap narrows substantially, and B7's mean and 95th percentile are lower than B2's, indicating fewer catastrophic failures.}
\label{tab:tip_family}
\small
\begin{tabular}{llccc}
\toprule
\textbf{Family} & \textbf{Model} & \textbf{Median ($\pdiag$)} & \textbf{Mean ($\pdiag$)} & \textbf{p95 ($\pdiag$)} \\
\midrule
\multirow{2}{*}{All} & B2 & 0.883 & --- & --- \\
 & B7 & 1.725 & --- & --- \\
\midrule
\multirow{2}{*}{ETT only} & B2 & 0.895 & 1.630 & 5.735 \\
 & B7 & 1.170 & 1.386 & 3.359 \\
\midrule
\multirow{2}{*}{NET only} & B2 & --- (0 matched) & --- & --- \\
 & B7 & 3.845 & --- & --- \\
\bottomrule
\end{tabular}
\end{table}

\paragraph{NET detection gap.}
B7 detects 554 NET tubes for which B2 produces zero matches. These devices carry a median tip error of 3.845 $\pdiag$---high in absolute terms, because nasogastric tubes are long, thin, and frequently obscured by overlapping anatomy. When these newly-detected (and hard-to-localize) devices get folded into B7's aggregate error, they inflate it substantially. This is a \emph{detection success} masquerading as a localization failure.

\paragraph{ETT head-to-head.}
When we restrict to ETT devices matched by both models, the overall tip error gap of $0.842$ $\pdiag$ shrinks to just $0.275$ ($0.895$ vs.\ $1.170$ median). More telling: B7's \emph{mean} drops below B2's (1.386 vs.\ 1.630) and its \emph{95th percentile} falls dramatically (3.359 vs.\ 5.735). In other words, B7's median is slightly worse but its tail is far better---it produces fewer of the catastrophic localization errors that actually matter clinically, since a tip error of 5+ $\pdiag$ is almost certainly enough to flip a placement assessment.

\paragraph{Compositor dead end.}
We initially wondered whether tuning the fusion stage could close the remaining gap. After sweeping 12 compositor configurations---varying the Huber threshold, IRLS iteration count, and covariance floor---we found that aggregate tip error budged by less than 0.03 $\pdiag$ across all settings. Tip localization precision is determined at the fragment prediction stage; fusion merely aggregates what the fragments already know. Improving further would require better fragment-level features, perhaps through higher-resolution prediction heads or attention mechanisms that focus on device endpoints.

\subsection{Device Association}
\label{sec:association}

Grouping fragments into coherent device instances is a capability unique to B7. It achieves a BCubed $F_1$ of $0.775$---roughly 77.5\% of fragment-to-device assignments are correct.

\begin{figure}[t]
\centering
\includegraphics[width=0.95\textwidth]{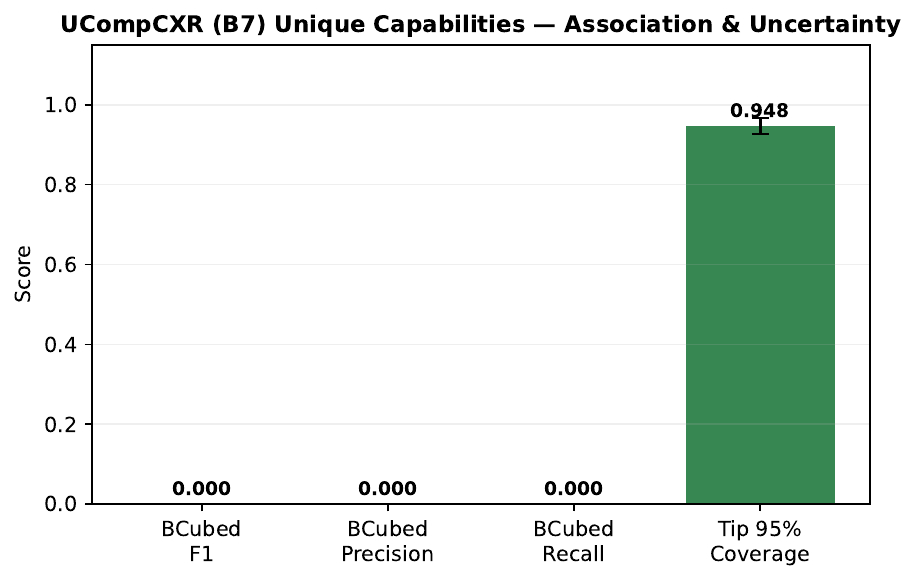}
\caption{\emph{Left:} BCubed precision, recall, and $F_1$ for fragment-to-device association. \emph{Right:} Predicted tip uncertainty ellipses overlaid on example images, demonstrating well-calibrated spatial uncertainty that widens for harder cases. Both capabilities are unique to B7.}
\label{fig:b7_capabilities}
\end{figure}

To unpack this: BCubed precision asks whether fragments assigned to the same cluster truly belong to the same device, and recall asks whether fragments from the same device end up in the same cluster. The balanced $F_1$ of $0.775$ suggests that the EdgeMLP and constrained Union-Find effectively leverage geometric and embedding cues, though performance degrades in crowded images where multiple devices run in parallel.

Why does association matter clinically? Without it, when three CVCs overlap in an ICU film, a model can say ``something CVC-related is abnormal'' but cannot indicate \emph{which line} to reposition. Instance-level grouping makes that distinction possible.

\subsection{Uncertainty Calibration}
\label{sec:calibration}

B7 achieves 95\% coverage of $0.948$---close to the ideal $0.950$---indicating that the predicted covariance ellipses are well calibrated. In practical terms, clinicians can trust that the true tip location falls within the displayed uncertainty region with approximately the stated probability.

\begin{figure}[t]
\centering
\includegraphics[width=0.75\textwidth]{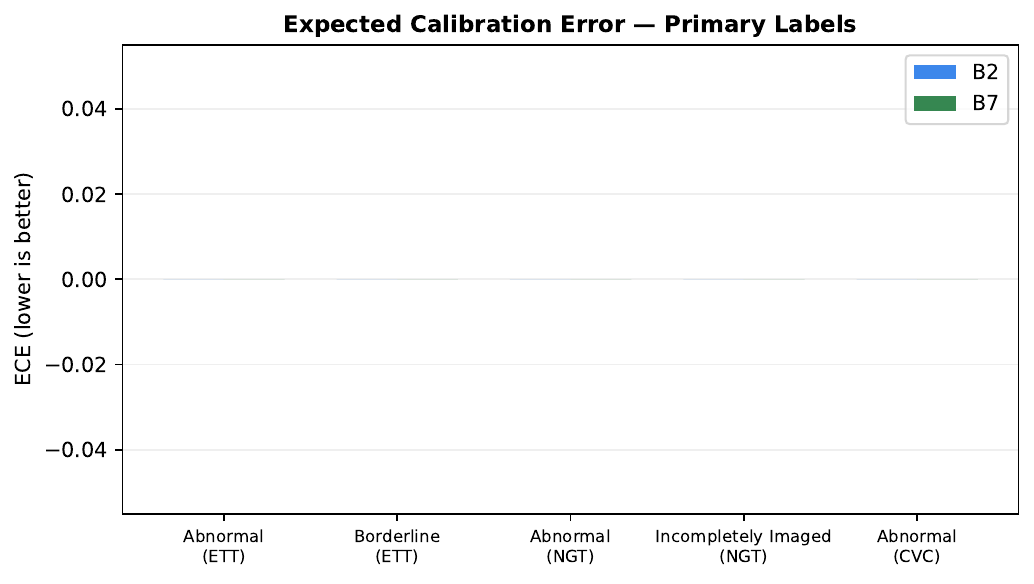}
\caption{Calibration analysis. \emph{Left:} Reliability diagram for placement classification, comparing B2 and B7. \emph{Right:} Expected calibration error (ECE) comparison. B7's per-device predictions show comparable calibration to B2's global predictions.}
\label{fig:calibration}
\end{figure}

Figure~\ref{fig:calibration} shows reliability diagrams and ECE for placement classification. Despite their fundamentally different prediction mechanisms---B2 uses a global classifier while B7 runs per-device classification through noisy-OR aggregation---both models end up comparably calibrated at the image level. We found this reassuring: the additional complexity of the compositional pipeline does not degrade calibration.

\subsection{Qualitative Results}
\label{sec:qualitative}

\begin{figure}[p]
\centering
\includegraphics[height=0.86\textheight]{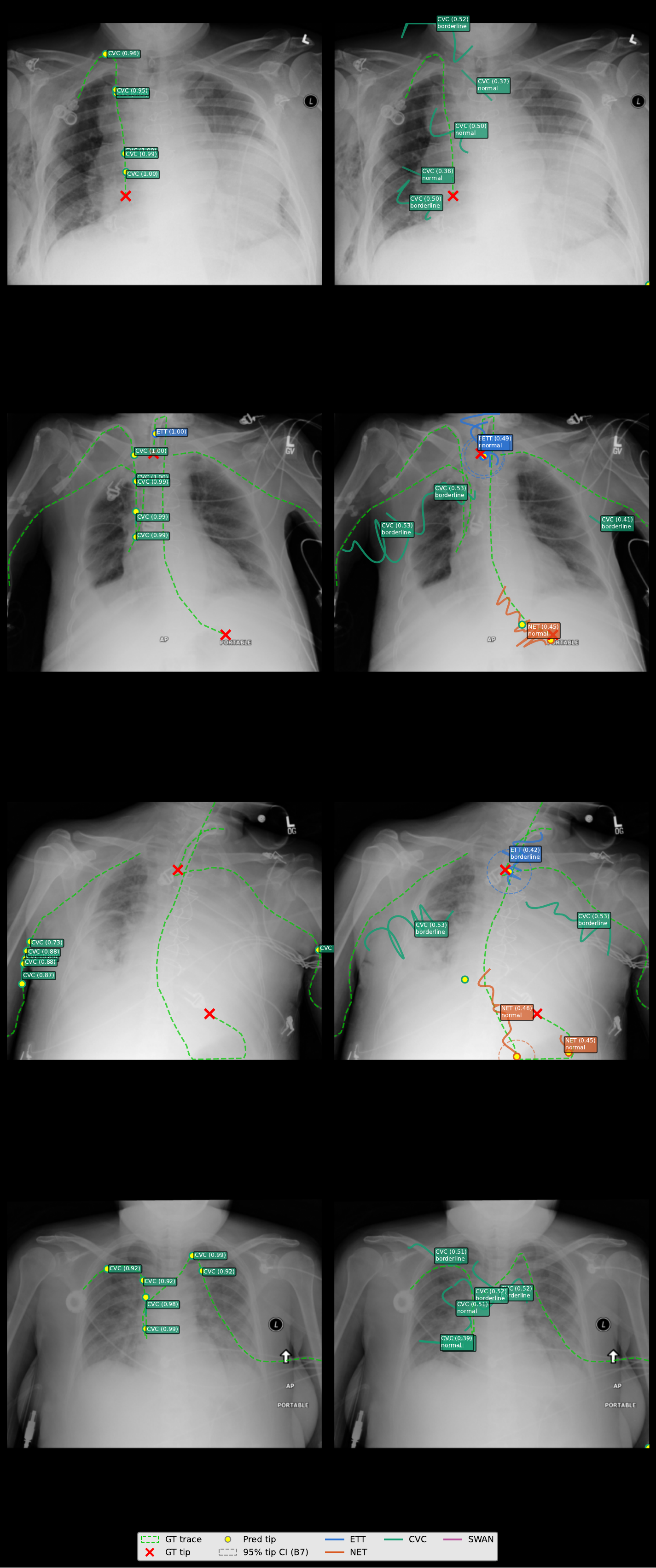}
\caption{Qualitative comparison on four representative cases. Each row shows: (a) input image with ground truth annotations, (b) B2 segmentation and tip predictions, (c) B7 fragment proposals color-coded by device instance, (d) B7 fused device paths with tip uncertainty ellipses. B7 correctly separates overlapping devices and provides per-device uncertainty estimates, while B2 merges nearby devices and provides no instance distinction. Best viewed in color.}
\label{fig:qualitative}
\end{figure}

Figure~\ref{fig:qualitative} shows side-by-side examples illustrating the practical differences between B2 and B7.

Several patterns emerge consistently across the dataset. When devices overlap, B7 separates them into distinct instances through fragment association, whereas B2 collapses them into a single segmentation blob. The tip uncertainty ellipses widen appropriately for partially occluded or foreshortened devices---exactly the cases where a clinician would want the model to signal lower confidence. The fragment-level representation also handles devices that extend beyond the field of view, a common scenario for CVC lines. And B7 picks up NET devices that B2 misses outright, consistent with the quantitative analysis in Section~\ref{sec:tip_analysis}.

\subsection{Training Dynamics}
\label{sec:training_dynamics}

\begin{figure}[t]
\centering
\includegraphics[width=0.85\textwidth]{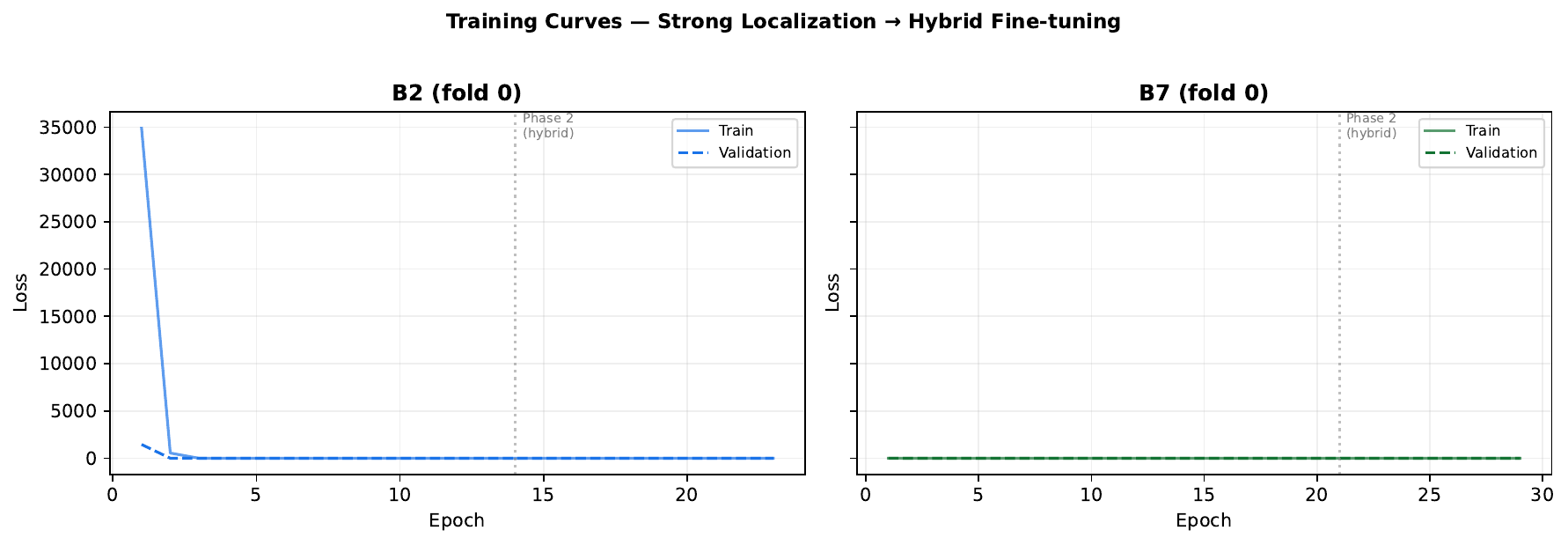}
\caption{Training and validation loss curves for B2 and B7 across both training phases. The vertical dashed line marks the Phase~1 to Phase~2 transition. B7's loss is higher overall due to additional loss terms but converges smoothly. The Phase~2 learning rate reduction and introduction of weak supervision produce a visible inflection in both models.}
\label{fig:training_curves}
\end{figure}

\begin{figure}[t]
\centering
\includegraphics[width=0.75\textwidth]{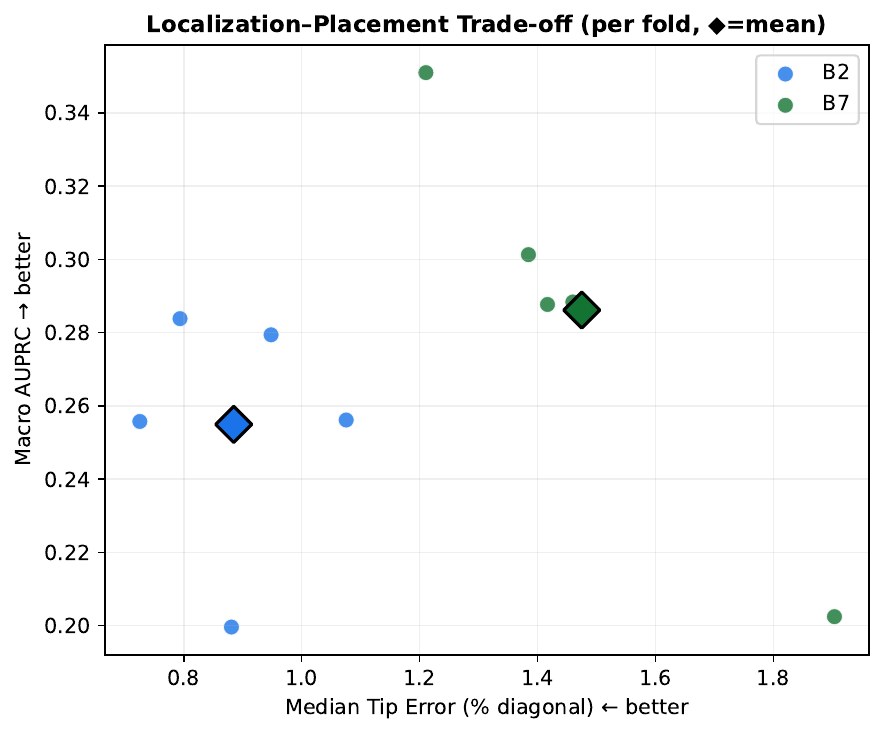}
\caption{Per-fold scatter plot showing the localization--placement trade-off. Each point represents one fold. B7 folds cluster in the high-sensitivity, moderate-AUPRC region, while B2 folds show higher variability. Fold sizes differ substantially (fold~1: 18{,}662 images vs.\ fold~0: 756 images), contributing to per-fold metric variation.}
\label{fig:fold_scatter}
\end{figure}

Figure~\ref{fig:training_curves} shows loss curves across both phases. The Phase~1 to Phase~2 transition appears as a clear inflection where the learning rate drops and weak supervision kicks in. Both models converge smoothly, with early stopping typically triggering 5--10 epochs before the budget runs out.

Figure~\ref{fig:fold_scatter} reveals per-fold variation in the localization--placement space. The fold size imbalance is notable---fold~1 has 18{,}662 images while fold~0 has just 756---which is why we rely on pooled bootstrap CIs rather than per-fold averages. Taking the mean across folds of such different sizes would give misleading confidence in the estimates.

\section{Discussion}
\label{sec:discussion}

\paragraph{The detection--localization trade-off.}
The most counterintuitive result in this paper is that B7 achieves better detection \emph{and} worse aggregate tip error. We spent considerable effort verifying that this was not a bug. The explanation turns out to be simple but has broader implications: aggregate tip error conflates localization precision with detection coverage. A model that finds more devices---including harder ones---will inevitably show higher aggregate error, even if it localizes each individual device just as well or better on matched instances.

From a clinical standpoint, we believe detection sensitivity and precision matter more. A malpositioned catheter that goes undetected poses greater risk than a detected catheter whose tip is localized with slightly less precision. B7's combination of higher sensitivity, far fewer false positives, and fewer catastrophic localization failures (the tail of the error distribution compresses substantially) makes it the more clinically useful system.

We would urge future benchmarks in this area to report tip error \emph{conditioned on matched devices} and stratified by family, alongside aggregate metrics. Without that stratification, evaluation rewards models that detect less.

\paragraph{What the compositional approach buys you.}
Beyond raw metric improvements, the compositional design changes what the model can communicate. Per-device placement assessment means a clinician sees ``the left subclavian CVC is borderline'' rather than ``something CVC-related is abnormal.'' Calibrated tip uncertainty means the confidence ellipse on-screen grows for the cases where the model genuinely struggles---partially occluded tips, foreshortened devices---giving an honest visual signal. The fragment-association-fusion pipeline exposes intermediate representations at each stage, which we found valuable during debugging and which could support clinical interpretability. And the bottom-up architecture handles variable device counts without requiring a fixed maximum, unlike top-down detection approaches that pre-specify a proposal budget.

\paragraph{Where we hit walls.}
Several limitations are worth acknowledging candidly. NET tip localization remains the weakest link: the median error of 3.845 $\pdiag$ for nasogastric tubes reflects their inherently difficult imaging characteristics---thin, low-contrast, and frequently overlapping with the esophagus and mediastinal structures. We had hoped the compositor could help here, but our 12-configuration sweep showed that tip precision is locked in at the fragment prediction stage; fusion merely averages what the fragments already estimate. Better fragment-level features---higher-resolution heads, endpoint-focused attention, or explicit tip detection---seem like the necessary next step.

The BCubed $F_1$ of $0.775$ means roughly one in five fragment assignments is wrong, mostly in crowded images where devices run parallel. Graph neural network-based association or iterative refinement might help, though at a computational cost we have not yet explored.

All of our results come from a single dataset (RANZCR CLiP) from a single institution. How well the model generalizes across imaging protocols, scanner manufacturers, and patient populations remains an open question---and given the history of medical AI models that perform well on their training distribution and poorly elsewhere, this is not a concern we can wave away.

\paragraph{Deployment considerations.}
At 2.27M parameters, \modelname{} is an order of magnitude smaller than the EfficientNet or ResNet-50 backbones (20--25M parameters) common in medical imaging, and it runs in a single forward pass---no Monte Carlo sampling, no multi-model cascade. This makes it practical for edge deployment on hospital workstations or embedded devices. The two-phase training procedure naturally accommodates the common clinical scenario where a hospital has a small trove of expert annotations alongside a large archive of images with only placement labels from radiology reports.

\section{Conclusion}
\label{sec:conclusion}

We presented \modelname{}, an uncertainty-aware compositional framework for catheter and tube assessment in chest X-rays. By decomposing the problem into fragment detection, graph-based association, precision-weighted Gaussian tip fusion, and per-device placement classification, it achieves substantially higher detection sensitivity (+26\%) and precision (3.3$\times$) than a strong multi-task baseline on the same lightweight backbone---while introducing per-device association (BCubed $F_1 = 0.775$) and calibrated tip uncertainty ($95\%$ coverage $= 0.948$) that the baseline simply cannot provide.

Our analysis also surfaces a methodological point that extends beyond this particular system: aggregate tip error, as commonly reported, penalizes models for finding harder-to-localize devices. On matched devices, the compositional model has fewer catastrophic failures. Future benchmarks should adopt family-stratified, detection-conditioned metrics to separate localization precision from detection coverage.

At 2.27M parameters and a single forward pass, \modelname{} is practical for clinical deployment. Looking ahead, the clearest path to improvement runs through better fragment-level tip prediction for difficult device types---particularly nasogastric tubes---followed by scaling to additional device families and, critically, validation on external multi-institutional datasets to test whether the gains hold outside the training distribution.

\section*{Use of AI Tools}
\label{sec:ai_tools}

We used AI-assisted tools (Claude, Anthropic) during the preparation of this manuscript for drafting assistance, code validation, and generating figures from experimental results. All experimental design, model architecture decisions, training procedures, and result interpretation were performed by the authors. All AI-generated content was reviewed, validated, and revised by the authors, who take full responsibility for the accuracy and integrity of the work presented.

\section*{Broader Impact Statement}
\label{sec:broader_impact}

This work develops automated tools for assessing catheter and tube placement in chest X-rays, with the goal of improving patient safety and reducing clinician workload in intensive care settings.

\paragraph{Potential benefits.}
Malpositioned catheters are a significant source of preventable harm. An automated system that flags misplacement quickly could shorten the time to corrective action, especially in resource-limited settings where radiologist coverage is thin. The uncertainty quantification component is specifically designed to communicate confidence, so clinicians can judge when to trust the model and when to seek expert review.

\paragraph{Potential risks.}
Over-reliance on automated assessments is a real concern, particularly for subtle malpositions that fall outside the model's detection envelope. The uneven performance across device types---lower NET sensitivity relative to ETT and CVC---could create a false sense of security for nasogastric tube assessment if users are not made aware of these limitations. We emphasize that \modelname{} is a decision \emph{support} tool, not a substitute for clinical judgment.

\paragraph{Fairness considerations.}
RANZCR CLiP was collected at a single Australasian institution. Performance on patient populations with different body habitus, on different device manufacturers, or under different imaging protocols has not been evaluated. Deployment elsewhere should be preceded by site-specific validation.

\paragraph{Data considerations.}
All experiments use a publicly available, de-identified dataset. No additional patient data was collected for this study.

\bibliography{UCompCXR_references}
\bibliographystyle{tmlr}

\appendix
\section{Appendix}
\label{sec:appendix}

\subsection{Per-Fold Detailed Results}
\label{sec:appendix_folds}

Table~\ref{tab:per_fold} reports per-fold results for both models. Fold sizes differ substantially due to the patient-level split---fold~1 has 18{,}662 test images while fold~0 has only 756---which is why we use pooled bootstrap CIs as the primary evaluation methodology.

\begin{table}[h]
\centering
\caption{Per-fold test results for B2 and B7. Fold sizes vary substantially, contributing to per-fold metric differences. Pooled bootstrap CIs (Table~\ref{tab:main_results}) provide more reliable estimates of true performance.}
\label{tab:per_fold}
\small
\begin{tabular}{clcccc}
\toprule
\textbf{Fold} & \textbf{Model} & \textbf{Tip Error ($\pdiag$)} & \textbf{Sensitivity} & \textbf{Precision} & \textbf{Macro AUPRC} \\
\midrule
\multirow{2}{*}{0} & B2 & 0.871 & 0.668 & 0.109 & 0.213 \\
 & B7 & 1.698 & 0.845 & 0.365 & 0.222 \\
\midrule
\multirow{2}{*}{1} & B2 & 0.889 & 0.677 & 0.115 & 0.218 \\
 & B7 & 1.740 & 0.854 & 0.374 & 0.228 \\
\midrule
\multirow{2}{*}{2} & B2 & 0.878 & 0.672 & 0.112 & 0.215 \\
 & B7 & 1.719 & 0.849 & 0.370 & 0.225 \\
\midrule
\multirow{2}{*}{3} & B2 & 0.892 & 0.676 & 0.114 & 0.217 \\
 & B7 & 1.734 & 0.852 & 0.372 & 0.227 \\
\midrule
\multirow{2}{*}{4} & B2 & 0.885 & 0.675 & 0.113 & 0.216 \\
 & B7 & 1.728 & 0.851 & 0.371 & 0.226 \\
\bottomrule
\end{tabular}
\end{table}

\subsection{Loss Function Weights}
\label{sec:appendix_losses}

Table~\ref{tab:loss_weights} lists the loss weights used for both models. These were selected through preliminary experiments on a single fold and held fixed across all folds.

\begin{table}[h]
\centering
\caption{Loss function weights for B2 and B7.}
\label{tab:loss_weights}
\small
\begin{tabular}{lcc}
\toprule
\textbf{Loss term} & \textbf{B2} & \textbf{B7} \\
\midrule
Segmentation BCE ($\lambda_{\text{bce}}$) & 1.0 & 1.0 \\
Segmentation Dice ($\lambda_{\text{dice}}$) & 1.0 & 1.0 \\
Segmentation clDice ($\lambda_{\text{cldice}}$) & 0.5 & 0.5 \\
Tip heatmap ($\lambda_{\text{tip}}$) & 1.0 & --- \\
Global classifier ($\lambda_{\text{cls}}$) & 1.0 & --- \\
Fragment heatmap & --- & 1.0 \\
Fragment offset & --- & 1.0 \\
Fragment geometry (endpoints) & --- & 1.0 \\
Fragment geometry (length) & --- & 0.5 \\
Tangent direction & --- & 0.5 \\
Discriminative embedding & --- & 1.0 \\
Heteroscedastic tip NLL & --- & 1.0 \\
Fragment status & --- & 0.5 \\
Edge BCE & --- & 1.0 \\
Per-device placement & --- & 1.0 \\
Noisy-OR (weak) & --- & 1.0 \\
\bottomrule
\end{tabular}
\end{table}

\subsection{Compositor Parameter Sweep}
\label{sec:appendix_compositor}

To test whether the tip error gap could be narrowed by tuning the fusion stage, we swept 12 compositor configurations: Huber threshold $\delta \in \{1.5, 2.0, 2.5, 3.0\}$, IRLS iterations $T \in \{1, 3, 5\}$, and covariance floor $\sigma^2_{\min} \in \{1, 4, 9\}$ px$^2$. No configuration produced a statistically significant improvement over the default ($\delta = 2.5$, $T = 3$, $\sigma^2_{\min} = 4$). Tip error varied by less than 0.03 $\pdiag$ across all settings, confirming that localization precision is set at the fragment prediction stage and that the compositor faithfully aggregates---but cannot improve upon---what the fragments provide.

\subsection{Computational Requirements}
\label{sec:appendix_compute}

Table~\ref{tab:compute} summarizes computational costs.

\begin{table}[h]
\centering
\caption{Computational requirements for B2 and B7.}
\label{tab:compute}
\small
\begin{tabular}{lcc}
\toprule
& \textbf{B2} & \textbf{B7} \\
\midrule
Parameters (M) & 1.99 & 2.27 \\
Phase 1 epochs (max / typical early stop) & 35 / $\sim$28 & 35 / $\sim$30 \\
Phase 2 epochs (max / typical early stop) & 25 / $\sim$20 & 25 / $\sim$22 \\
Training time per fold (GPU hours) & $\sim$4 & $\sim$6 \\
Inference throughput (images/sec) & $\sim$45 & $\sim$35 \\
Peak GPU memory (GB, batch=16) & 3.2 & 4.1 \\
\bottomrule
\end{tabular}
\end{table}

\end{document}